\documentclass[a4paper]{article}

\usepackage{INTERSPEECH2021}
\usepackage{multirow, cite}
\usepackage{threeparttable}
\usepackage{enumitem}
\usepackage{hyperref}

\title{MFA-Conformer: Multi-scale Feature Aggregation Conformer for \\ Automatic Speaker Verification}
\name{Yang Zhang$^{1,2}$, Zhiqiang Lv$^2$, Haibin Wu$^{3,4}$, Shanshan Zhang$^2$, Pengfei Hu$^2$, \\ 
Zhiyong Wu$^{1,4,\dagger}$, Hung-yi Lee$^3$,
Helen Meng$^4$
\thanks{$\dagger$ Corresponding author. } 
}

\address{
  $^1$ Shenzhen International Graduate School, Tsinghua University\\
  $^2$ TEG AI, Tencent Inc\\
  $^3$ Graduate Institute of Communication Engineering, National Taiwan University \\
  $^4$ Centre for Perceptual and Interactive Intelligence, The Chinese University of Hong Kong
}
\email{zhangy20@mails.tsinghua.edu.cn, zywu@sz.tsinghua.edu.cn}

\begin{document}

\maketitle
\begin{abstract}
In this paper, we present Multi-scale Feature Aggregation Conformer (MFA-Conformer), an easy-to-implement, simple but effective backbone for automatic speaker verification based on the Convolution-augmented Transformer (Conformer).
The architecture of the MFA-Conformer is inspired by recent state-of-the-art models in speech recognition and speaker verification.
Firstly, we introduce a convolution subsampling layer to decrease the computational cost of the model.
Secondly, we adopt Conformer blocks which combine Transformers and convolution neural networks (CNNs) to capture global and local features effectively.
Finally, the output feature maps from all Conformer blocks are concatenated to aggregate multi-scale representations before final pooling. 
We evaluate the MFA-Conformer on the widely used benchmarks. 
The best system obtains 0.64\%, 1.29\% and 1.63\% EER on VoxCeleb1-O, SITW.Dev, and SITW.Eval set, respectively.
MFA-Conformer significantly outperforms the popular ECAPA-TDNN systems in both recognition performance and inference speed.
Last but not the least, the ablation studies clearly demonstrate that the combination of global and local feature learning can lead to robust and accurate speaker embedding extraction.
We have also released the code\footnote{\url{https://github.com/zyzisyz/mfa_conformer}} for future comparison. 

\end{abstract}

\section{Introduction}



Automatic speaker verification (ASV) is a task to verify whether a given utterance is from a claimed enrolled speaker. 
In recent years, it is witnessed the significant developments of ASV\cite{dehak2010front, kenny2007joint, variani2014deep,snyder2018x,chung2020defence,desplanques2020ecapa} and now ASV is a well-developed technology and widely employed in real-world applications, such as intelligent housing systems, law enforcement and real-time online meeting.
Modern speaker verification technologies are mainly based on \emph{deep speaker embedding} approach, which maps a piece of variable length speech to a fixed-dimension embedding through deep neural networks.
The speaker embeddings extracted by ASV also serve as a key component for speaker diarization, voice conversion / cloning and speech recognition.

Convolution neural networks (CNNs) based models have achieved a remarkable success for ASV. 
X-vector\cite{snyder2018x} firstly employs the time delay neural networks (TDNNs) to maps variable-length utterances to fixed-dimensional embeddings.
Later, 
equipped with residual connections,  ResNet-based\cite{he2016deep,zeinali2019but} r-vector system and its variantions\cite{jung2019RawNet,chung2020defence,zhou2021resnext} become capable of training deeper networks and have shown a outstanding results.
Recently, 
based on blocks of TDNNs and squeeze and excitation (SE)\cite{hu2018squeeze} layers unified with Res2Block\cite{gao2019res2net},
the ECAPA-TDNN\cite{desplanques2020ecapa} and its subsequent efforts\cite{thienpondt2021integrating,liu2022mfa} achieve a significant breakthrough and deliver the equal error rates less than 1\% in VoxCeleb1-O benchmark.

Despite the great success, CNN still has its limitations. It mainly focuses on local spatial modeling, but lacks of global context fusion.
CNNs-based models can not handle the long-range dependencies very well.
To overcome this issue, Transformer\cite{vaswani2017attention} and its variations\cite{dai2019transformer,yang2019xlnet,zhou2021informer} have become a prevalent architecture in many sequence processing tasks.
Transformers are good at modeling long-range global context and facilitate efficient parallel training.
However, many existing studies such as UniSpeech\cite{wang2021unispeech} and WavLM\cite{chen2021wavlm,chen2021large}, indicate that without complicated pre-training procedures and large parameters, Transformer can hardly achieve a satisfactory performance in ASV.
Recently, 
Convolution-augmented Transformer (Conformer)\cite{gulati20_interspeech}, a combination of CNNs and Transformers, becomes a promising candidate for advancing speech processing performance.
Conformer inserts convolution modules into Transformer to increase the local information modeling. 
It firstly achieves an outstanding results in end-to-end speech recognition, and is later adopted in speech enhancement\cite{koizumi2021df} and speech separation\cite{chen2021continuous} with remarkable performance. 

Inspired by these recent progresses,
we proposed MFA-Conformer, an easy-to-implement and effective backbone for speaker embedding extraction.
Firstly, the input acoustic feature is processed by a convolution subsampling layer to decrease the computational cost.
Secondly, we adopt Conformer blocks which combine Transformers and convolution neural networks to capture the global and local features effectively.
Finally, we concatenate the output features from all Conformer blocks to aggregate the multi-scale representations before final pooling.
Our contributions can be summarized as follow:

\begin{figure*}[htb]
\centering 
\includegraphics[width=17cm]{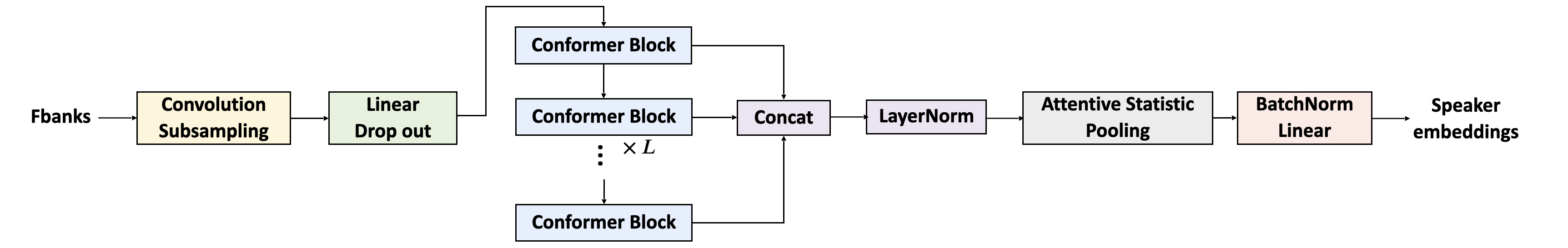} 
\caption{The overall architecture of Multi-scale Feature Aggregation Conformer (MFA-Conformer)} 
\label{conformer} 
\end{figure*}

\begin{itemize}[itemsep=0pt,topsep=0pt,parsep=0pt,leftmargin=10pt]
    \item To our  best knowledge, this study is the first to adopt Conformer blocks with elaborated modification for ASV. This work reveals that the Transformer-based model can achieve a remarkable performance in ASV without any complicated pre-training procedures or large network parameters. 
    \item The proposed MFA-Conformer significantly outperforms the popular CNNs-based ECAPA-TDNN baseline systems in both inference speed and accuracy. Other than that, it achieves state-of-the-art results on SITW \cite{mclaren2016speakers} benchmark.
    \item We conduct numerous experiments and the results demonstrate the combination of local and global modeling can lead to the robust speaker embedding extraction. Especially in the real-world where the speech lengths are different, the MFA-Conformer shows robust performance as mentioned in Section~\ref{sec:global feature modeling} 
\end{itemize}

\section{MFA-Conformer}
\label{sec3}

In this section, we illustrate the fundamental components of the proposed Multi-scale Feature Aggregation Conformer (MFA-Conformer) system.
The overall architecture is shown in Fig.\ref{conformer}.

\subsection{The Conformer block}



Local features and global dependencies are both crucial for speaker representation learning.
Local features, including the pitch, intonation style and pronunciation pattern, are essential to represent the speaker characteristics.
And the good capacity of global context modeling to capture the long-range dependencies of variable-length speech will lead to robust speaker embedding extraction.
Convolution neural networks are good at extracting local features
but is weak at capturing global representations.
Self-attention from Transformer can capture long-range global context dependencies but is lack of local details.

In order to both model the global and local features more directly and efficiently, 
we employ the Conformer block\cite{gulati20_interspeech}, a combination of CNNs and Transformers, to better capture the speaker characteristics. 
Multi-head self-attention (MHSA) and convolution modules are the key components for a Conformer block. 
The MHSA in the Conformer employs the relative positional encoding scheme proposed in Transformer-XL\cite{dai2019transformer}.
It encodes inputs according to the relative position deviation and takes account of both the global content offset and global position offset. 
Therefore, the relative positional encoding MHSA module is more robust for the utterances with variable input lengths.
The convolution module after the MHSA consists of Pointwise convolution, 1D Depthwise convolution and BatchNorm after the convolution layer contributing to training deep models easier.

The structure of the Conformer block\cite{gulati20_interspeech} is different from the Transformer block\cite{vaswani2017attention}. 
It contains two Macaron-like feed forward modules (FNN) with half residual connections sandwiching the MHSA and convolution modules (Conv).
Mathematically, for the input feature $\boldsymbol{h_{i-1}}$
to the $i$ Conformer block, the output feature $\boldsymbol{h_{i}}$ is calculated by:
\begin{equation}
\begin{aligned}
\widetilde{\boldsymbol{h}}_i &= \boldsymbol{h_{i-1}} + \frac{1}{2}{\rm FNN}(\boldsymbol{h_{i-1}}) \\
\boldsymbol{h'_i} & = \widetilde{\boldsymbol{h}}_i + {\rm MHSA}(\widetilde{\boldsymbol{h}}_i) \\
\boldsymbol{h''_i} &= \boldsymbol{h'_i} + {\rm Conv}(\boldsymbol{h'_i}) \\
\boldsymbol{h_{i}} &= {\rm LayerNorm}(\boldsymbol{h''_i}+\frac{1}{2}{\rm FNN}(\boldsymbol{h''_i}))
\end{aligned}
\end{equation}
Note that $\boldsymbol{h_{i-1}}\in \mathbb{R}^{d\times T}$ and $\boldsymbol{h_{i}}\in \mathbb{R}^{d\times T}$, where $d$ denotes the Conformer encoder dimension and $T$ denotes the frame length.


\subsection{MFA with attentive statistics pooling}
 
Previous studies\cite{gao2019improving,tang2019deep} indicate that the low-level feature maps can also contribute towards the accurate speaker embedding extraction. 
Based on this experience, in ECAPA-TDNN system, the output feature maps from all SE-Res2blocks are aggregated before the final pooling layer, and this aggregation leads to an obvious performance improvement.
Likewise,
we concatenate the output feature maps from each Conformer block and then feed them into a LayerNorm layer:
\begin{equation}
\begin{aligned}
\boldsymbol{H'} &= {\rm Concat}(\boldsymbol{h_1}, \boldsymbol{h_2}, ..., \boldsymbol{h_L}) \\
\boldsymbol{H} &= {\rm LayerNorm}(\boldsymbol{H'})
\end{aligned}
\end{equation}
where $\boldsymbol{H'} \in \mathbb{R}^{D \times T}$ and $\boldsymbol{H}=[H_1, H_2, ..., H_T] \in \mathbb{R}^{D \times T}$.
$L$ denotes the number of Conformer blocks and $D=d\times L$. 

Furthermore, we adopt the attentive statistics pooling 
\cite{okabe2018attentive,desplanques2020ecapa} to capture the importance of each frame and extract more robust speaker embedding.
Specifically, for a frame-level feature $H_t$ at time step $t$, we firstly calculate scalar score $e_t$ and normalized score $\alpha_t$ as:
\begin{equation}
\begin{aligned}
e_{t} &= \boldsymbol{v}^{T}f(\boldsymbol{W}H_t+\boldsymbol{b})+k \\
\alpha_t &= \frac{\exp(e_t)}{\sum_{\tau=1}^{T}\exp(e_\tau)}
\end{aligned}
\end{equation}
where $\boldsymbol{W}\in \mathbb{R}^{D \times D}$, $\boldsymbol{b}\in \mathbb{R}^{D\times 1}$, $\boldsymbol{v}\in \mathbb{R}^{D\times 1}$ and $k$ are the trainable 
parameters
for attention.
$f(\cdot)$ denotes the Tanh activation function.
After that, the normalized score $\alpha_t$ is adopted as the weight to calculate the weighted mean vector $\boldsymbol{\widetilde\mu}$ and weighted standard deviation $\boldsymbol{\widetilde\sigma}$, which are formulated as:
\begin{equation}
\begin{aligned}
\boldsymbol{\widetilde\mu} &= \sum_{t=1}^T \alpha_tH_t \\
\boldsymbol{\widetilde\sigma} &= \sqrt{\sum_{t=1}^T \alpha_tH_t \odot H_t-\mu \odot \mu}
\end{aligned}
\end{equation}
where $\mu=\frac{1}{T}\sum_{\tau=1}^TH_\tau$ and $\odot$ denotes the Hadamard product.
The output of the pooling layer is given by concatenating the vectors of the weighted mean $\boldsymbol{\widetilde\mu}$ and weighted standard deviation $\boldsymbol{\widetilde\sigma}$.

Finally, the speaker embedding is extracted from a high dimension vector to a low dimension vector with BatchNorm using the fully-connected linear layer.
\section{Experimental Setup}
\label{sec4}

In this section,
we present datasets, implementation details and evaluation protocols. 
In industrial applications scenario, such as video processing \& analysis, the speech lengths may vary from less
than 5 seconds to more than 30 seconds.
Therefore, VoxCeleb1-O and SITW benchmarks
are adopted to illustrate the advantages of MFA-Conformer in 
different utterance duration scenarios.


\subsection{Dataset}
\label{sec:data}

VoxCeleb1\&2\cite{Nagrani17,Chung18b} and SITW\cite{mclaren2016speakers} are used in our experiments. VoxCeleb is an audio-visual dataset consisting of 2,000+ hours short clips of human speech, extracted from interview videos on YouTube.
SITW is a widely-used standard evaluation dataset in real-world conditions, consisting of 299 speakers including two testing trials (SITW.Dev and SITW.Eval).
For model training, we use the development set of VoxCeleb1\&2, which contain 1,240,000+ utterances from 7,205 speakers.

To better illustrate the advantages of MFA-Conformer in different utterance duration conditions, we adopt 3 trials including Voxceleb1-O, SITW.Dev and SITW.Eval for recognition performance evaluation. 
Note that Voxceleb1-O is the test part of Voxceleb1 and
the major durations of testing utterances are 5-8s, which can be regarded as the short-duration utterance scenario.  
As for SITW, the major durations are about 30-40s, therefore it can be regarded as the long-duration scenario. 

\subsection{Network configurations}
\label{sec:net}

The speaker embedding dimension of all systems is 192.
For fair comparisons, we also re-implement the r-vector system proposed in \cite{zeinali2019but} and the ECAPA-TDNN system proposed in \cite{desplanques2020ecapa} as the baselines.
Other configurations are presented below:

\noindent \textbf{ResNet34.}
The first baseline is the ResNet-based r-vector system, 
which contains four residual blocks with different channels.
We set the channels of residual blocks as \{64, 128, 256, 512\}.
The total number of learnable parameters is 23.2M.

\noindent \textbf{ECAPA-TDNN.}
The second baseline is ECAPA-TDNN, which contains three carefully designed SE-Res2Blocks.
We set the channels of SE-Res2Blocks as \{1024, 1024, 1024\}.
The total number of learnable parameters is 20.8M.

\noindent \textbf{MFA-Conformer.}
The proposed MFA-Conformer, whose structure follows the practical experience in end-to-end speech recognition.
Specifically,
for multi-headed self-attention module,
we set the encoder dimension as 256 and set the number of attention heads as 4;
for convolution module, we set the kernel size to 15;
for feed forward module, we set linear hidden units as 2048.
We adopt 6 Conformer blocks with different subsampling rates. The total number of learnable parameters is about 19.7M-20.5M, which is closed to the above baseline systems.

\subsection{Implementation details}



We use Pytorch\cite{paszke2019pytorch} framework to implement the proposed MFA-Conformer and the baseline systems\footnote{Original code of the baseline systems is not available.
For a fair comparison, all the models are trained in the same framework. 
There may be a few mismatches between the re-implemented baseline systems and reference papers\cite{zeinali2019but,desplanques2020ecapa}. We borrow code from WeNet toolkit\cite{yao2021wenet} and the training details can be found in the accompanying open-sourced code.
}.
A fixed length 3-second segments are extracted randomly from each utterance. The input features are 80-dimensional Fbanks with a window length of 25 ms and a frame-shift of 10 ms. No voice activity detection or augmentation are performed.
All models are trained using additive margin Softmax (AM-Softmax) loss\cite{wang2018cosface} with a margin of 0.2 and a scaling factor of 30.
We use the Adam optimizer with an initial learning rate of 0.001 and decrease the learning rate by 50\% every 4 epoch. 
We also set the weight decay as 1e-7 to avoid overfitting and perform a linear warmup at the first 2k steps. The batch size is 200.

\subsection{System evaluation}
We use cosine distance with adaptive s-norm\cite{matejka2017analysis} for scoring. 
Then we report the Equal Error Rate (EER) and minimum Detection Cost Function (minDCF) with $P_{target} = 0.01$ and $C_{FA} = C_{Miss} = 1$ for performance evaluation.
Furthermore, we calculate the real-time factor (RTF) on an Intel(R) Xeon(R) Silver 4210R CPU (2.40GHz) to evaluate the inference speed.

\section{Experimental Results}
\label{sec5}

\subsection{Results on VoxCeleb test and SITW}

\begin{table*}[h]
\centering
\begin{threeparttable}
\caption{Performance overview of all systems on VoxCeleb1-O (short-duration), SITW.Dev \& SITW.Eval (long-duration)}
\begin{tabular}{lcccccccc}
\toprule[1pt]
\multicolumn{1}{c}{\multirow{2}{*}{\textbf{Model}}} & \multirow{2}{*}{\textbf{\# Parameters}} & \multirow{2}{*}{\textbf{RTF}} & \multicolumn{2}{c}{\textbf{VoxCeleb1-O}} & \multicolumn{2}{c}{\textbf{SITW.Dev}} & \multicolumn{2}{c}{\textbf{SITW.Eval}} \\
\multicolumn{1}{c}{}                                &                                     &                               & EER(\%)            & minDCF              & EER(\%)           & minDCF            & EER(\%)           & minDCF             \\ \hline
ResNet34                                            & 23.2M                               & 0.0088                        & 1.03               & 0.112               & 2.34              & 0.209             & 2.54              & 0.226              \\
ECAPA-TDNN                                          & 20.8M                               & 0.0180                        & 0.82               & 0.112               & 1.91              & 0.179             & 2.22              & 0.192              \\ \hline
MFA-Conformer (1/1)                                 & 20.5M                               & 0.0203                        & 0.83               & 0.102               & 1.51              & 0.159             & 1.78              & 0.172              \\
MFA-Conformer (1/2)                                 & 20.5M                               & 0.0121                        & \textbf{0.64}      & \textbf{0.081}      & \textbf{1.29}     & \textbf{0.137}    & \textbf{1.63}     & \textbf{0.153}     \\
MFA-Conformer (1/4)                                 & 19.8M                               & 0.0102                        & 0.83               & 0.118               & 1.88              & 0.160             & 1.94              & 0.178              \\
MFA-Conformer (1/6)                                 & 20.4M                               & 0.0093                        & 1.22               & 0.142               & 1.91              & 0.200             & 2.46              & 0.222              \\
MFA-Conformer (1/8)                                 & 19.7M                               & 0.0089                        & 1.48               & 0.182               & 2.88              & 0.240             & 2.79              & 0.261              \\ \toprule[1pt]

\end{tabular}
\begin{tablenotes}
\footnotesize
\item $^*$MFA-Conformer (1/2) means the convolution subsampling rate is 1/2.
\end{tablenotes}
\label{all}
\end{threeparttable}
\vspace{-10pt}
\end{table*}


In this section, we present the performance of the proposed MFA-Conformer with different subsampling rates, 
as well as the performance of the two baseline systems ResNet34 and ECAPA-TDNN. 
Table \ref{all} reports the equal error rate (EER) and minimum Detection Cost Function (minDCF) together with the number of model parameters and real time factor (RTF).
Firstly, it can be observed from the first and second lines that comparing with ResNet34, the ECAPA-TDNN system achieves an obvious advantage in recognition performance, yet the RTF is unsatisfied.
Secondly, from the results of the proposed MFA-Conformer with different subsampling rates, we find that MFA-Conformer (1/1), MFA-Conformer(1/2) and MFA-Conformer (1/4) achieve promising results.
Compared with the popular ECAPA-TDNN systems, the MFA-Conformer(1/2) obtains 21\% relative improvement in EER and 32\% relative improvement in inference speed (RTF).
Thirdly, as described in Section \ref{sec:data}, VoxCeleb1-O is a short-duration test scenario and SITW is a long-duration test scenario. We can find that the MFA-Conformer is able to obtain more competitive results than CNNs-based systems in long-duration utterance scenarios.
This also indicates that MFA-Conformer can better handle long-range sequences and extract a more reliable speaker embedding for long utterances.


\subsection{Impacts of global feature modeling}
\label{sec:global feature modeling}

\begin{figure}[h]
\centering
\includegraphics[width=8cm]{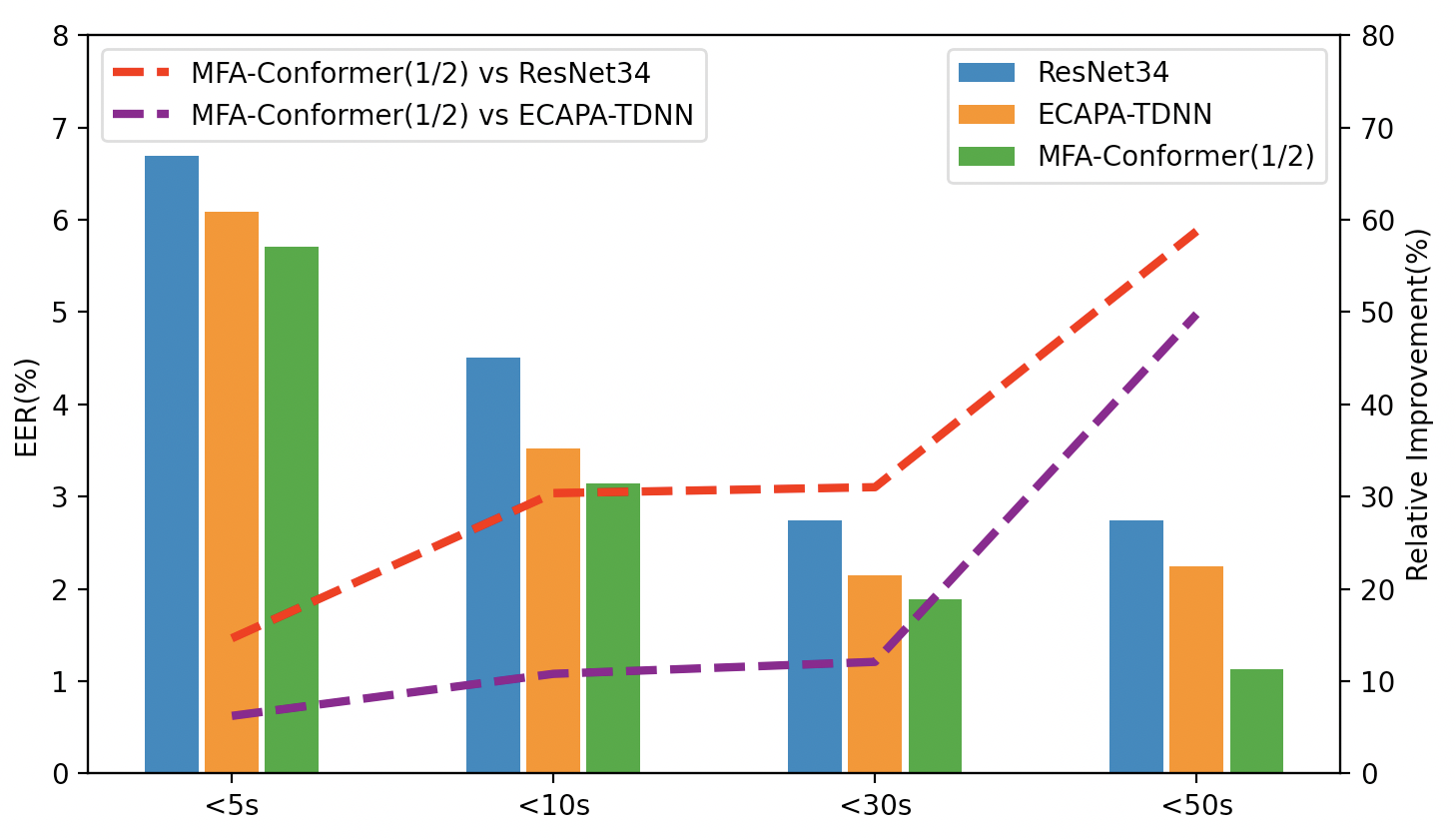}
\caption{Performance of MFA-Conformer (1/2) and two baselines with different utterance durations. The bars denote the EERs, and the dotted lines denote the relative improvement of MFA-Conformer (1/2) over the two baseline. }
\label{fig:duration}
\vspace{-15pt}
\end{figure}

In the real-world applications,
such as video processing, real-time online meeting,
the utterance lengths may vary from less than 5 seconds to more than 30 seconds.
Extracting robust global features for utterances with different lengths is important for speaker verification.
The Multi-head self-attention (MHSA) is the key component to make Transformer or Conformer unique from the widely used CNNs-based models.
In this section, we investigate the impacts of the global interactions modeling of MHSA by comparing the system performances with different utterance durations. 
We randomly split the test utterances of SITW according to different duration ranges to generate new trials,
and report the EERs of the new trials in Fig.\ref{fig:duration}. 
we can observe that when the test utterance duration becomes longer, the MFA-Conformer achieves larger relative improvements.
It further indicates that MFA-Conformer is better at modeling long-range, global context information.

\subsection{Impacts of local feature modeling}
\label{sec53}
The Conformer inserts an additional Convolution module into the Transformer to better capture the local information.
In this section, we study the impacts of the local feature modeling by comparing the performances between Transformer and Conformer blocks.
We replace the Conformer blocks by Transformer blocks and make a group of experiments with different number of blocks $L$ in SITW test set.
As shown in Table \ref{Transformer.Conformer}, it can be clearly observed that Conformer blocks have remarkable advantages over Transformer blocks.
MFA-Transformer can hardly obtain a satisfactory performance even increasing the number of the blocks.
This indicates that the local spatial modeling by convolution module plays a critical role in accurate speaker embedding extraction.
And the results show that setting $L$ to 6 performs better than the rests.

\begin{table}[h]
\caption{MFA-Conformer v.s. MFA-Transformer on SITW}
\centering
\setlength{\tabcolsep}{1.3mm}
\begin{tabular}{cccccc}
\toprule[1pt]
\multirow{2}{*}{\textbf{Block}}                                         & \multirow{2}{*}{\textbf{$L$}} & \multicolumn{2}{c}{\textbf{SITW.Dev}} & \multicolumn{2}{c}{\textbf{SITW.Eval}} \\
                                                                             &                               & EER(\%)           & minDCF            & EER(\%)           & minDCF             \\ \hline
\multirow{5}{*}{\begin{tabular}[c]{@{}c@{}}Transformer\\ Block\end{tabular}} & 1                             & 4.78              & 0.357             & 4.96              & 0.413              \\
                                                                             & 3                             & 3.47              & 0.272             & 3.31              & 0.303              \\
                                                                             & 6                             & 2.50              & 0.237             & 3.01              & 0.246              \\
                                                                             & 9                             & 2.56              & 0.221             & 2.48              & 0.232              \\
                                                                             & 12                            & 2.61              & 0.224             & 2.65              & 0.238              \\ \hline
\multirow{5}{*}{\begin{tabular}[c]{@{}c@{}}Conformer\\ Block\end{tabular}}   & 1                             & 3.38              & 0.280             & 3.73              & 0.308              \\
                                                                             & 3                             & 2.15              & 0.186             & 2.14              & 0.195              \\
                                                                             & 6                             & \textbf{1.29}     & \textbf{0.137}    & \textbf{1.63}     & 0.153              \\
                                                                             & 9                             & 1.45              & 0.141             & 1.76              & 0.158              \\
                                                                             & 12                            & 1.73              & 0.150             & 1.69              & \textbf{0.149}     \\ \toprule[1pt]
\end{tabular}
\label{Transformer.Conformer}
\end{table}

                                                                     

\subsection{Ablation Studies}
\label{sec54}

In the final section, 
we remove the individual components introduced in Section \ref{sec3} to study the effect of each component contributing to performance improvements.
Due to the space limitation, we only present the results on VoxCeleb1-O shown in Table \ref{abs}, and the results in other sets attain the same trend.
The results in the first line are the MFA-Conformer with 1/2 subsampling rate. 
In the second line, we remove relative positional encoding scheme;
in the third line, we remove the Macaron-style feed forward network and only keep the single feed-forward network after MHSA;
in the fourth line, we discard the multi-scale feature aggregation strategy and only use the output representations from the last Conformer block;
in the last line, we remove the convolution module to further measure the impact of the local feature modeling.
It can be observed that multi-scale feature aggregation and convolution module play the most critical roles in achieving the promising performance
Aggregation of the outputs from all blocks brings 48.3\% relative improvement in EER.
And convolution module leads to 54.9\% relative improvement in EER.

\begin{table}[h]
\centering
\caption{Ablation study of MFA-Conformer on VoxCeleb1-O.}
\begin{threeparttable}
\begin{tabular}{lcc}
\toprule[1pt]
\multicolumn{1}{c}{} & \textbf{EER(\%)} & \textbf{minDCF} \\ \hline
MFA-Conformer (1/2)        & \textbf{0.64}             & \textbf{0.081}           \\ \hline
w/o Relative PE      & 0.77             & 0.086           \\
w/o Macaron FFN      & 0.84             & 0.085           \\
w/o MFA              & 1.24             & 0.150           \\
w/o Conv             & 1.42             & 0.147           \\ \toprule[1pt]
\end{tabular}
\begin{tablenotes}
\footnotesize
\item $^*$w/o is without.
\end{tablenotes}
\end{threeparttable}
\label{abs}
\end{table}

\section{Conclusions}
\label{sec6}

This paper proposes MFA-Conformer, a novel backbone for automatic speaker verification.
MFA-Conformer could be an ideal backbone for real industry speaker recognition scenarios.
It significantly outperforms the popular ECAPA-TDNN systems in both recognition performance and inference speed. 
And it can extract more robust embeddings when the utterances are with different durations.
Our ablation study shows the combination of local and global feature modeling can lead to the robust speaker embedding extraction, this can provide inspiration for the future ASV system design and acceleration.
In the future work, we will extend the MFA-Conformer for streaming speaker recognition scenarios.

\section{Acknowledgements}
This work was supported by National Natural Science Foundation of China (62076144), CCF-Tencent Open Research Fund (RAGR20210122) and Shenzhen Science and Technology Innovation Committee (WDZC20200818121348001).

\bibliographystyle{IEEEtran}

\bibliography{ref.bib}

\end{document}